\documentclass[prl,twocolumn,amssymb]{revtex4}
\usepackage{epsfig}
\begin{document}
\title[]{Magnetothermal Conductivity
of Highly Oriented Pyrolytic Graphite in the Quantum Limit}
\date{\today}
\author{R. Oca\~na}
\author{P. Esquinazi}
\email[E-mail address: ]{esquin@physik.uni-leipzig.de}\author{H. Kempa}
\affiliation{Abteilung Supraleitung und Magnetismus, Institut f\"ur
Experimentelle Physik II, Universit\"at Leipzig, Linn{\'e}str. 5, D-04103
Leipzig, Germany} \author{J. H. S. Torres} \author{Y. Kopelevich}
\affiliation{Instituto de F{\'i}sica, Universidade Estadual de Campinas,
Unicamp 13083-970, Campinas, S\~{a}o Paulo, Brasil}

\begin{abstract}
We report on the magnetic field (0T$ \le B \le 9$T) dependence of the
longitudinal thermal conductivity $\kappa(T,B)$ of highly oriented
pyrolytic graphite in the temperature range 5~K $\le T\le$ 20~K for fields
parallel to the $c-$axis. We show that $\kappa(T,B)$ shows large
oscillations in the high-field region $(B > 2~$T) where clear signs of the
Quantum-Hall effect are observed in the Hall resistance. With the measured
longitudinal electrical resistivity we show that the Wiedemann-Franz law
is violated in the high-field regime.

 \end{abstract}
  \maketitle
\section{Introduction}

Over the last 60 years the literature contains a substantial
number of measurements and theoretical work on the electrical and
thermal transport properties of the semi-metal
graphite\cite{kel,dre}. In contrast to the common believe,
however, several transport properties of graphite are not
understood and some theoretical assumptions done in the past seem
now less plausible. These doubts have their origin in the
relatively new and controversial physics of a two-dimensional (2D)
electron gas. Graphite has been tacitly assumed to be a 2D
material, but the relatively low quality of the samples prevented
clear measurements of its true 2D transport properties.  The
number of open questions regarding the transport properties of
graphite is significant. Already early work noted that the
magnetic field dependence of the electrical longitudinal
resistivity $\rho_{xx}$ and Hall resistance $R_H$, even including
ad-hoc dispersion relation and trigonal warping of the constant
energy surfaces, was (and still remains) basically unexplained
\cite{dil}.

Recent measurements of the longitudinal resistivity of highly
oriented pyrolytic graphite (HOPG) samples show a clear magnetic
field driven metal-insulator transition (MIT) with a giant
magnetoresistance at low fields and at low temperatures
\cite{kempa1}. For example, the longitudinal resistivity at 4.2~K
can increase by more than one order of magnitude with a field $B
\simeq 0.2$~T. This MIT shows a scaling as found for 2D electron
systems with similar scaling exponent but with a critical field
$\sim 0.1~$T applied perpendicular to the graphene
layers\cite{kempa2,yakovadv}. Possible origins for the MIT in
graphite are being discussed nowadays in the literature in terms
of superconducting fluctuations \cite{kempa1}, excitonic insulator
state triggered by a magnetic catalysis phenomenon \cite{khv}
and/or a Bose metal transition \cite{das,yakovadv}.
High-resolution angle dependence of the magnetoresistance along
the $c-$axis in HOPG samples indicates that the transport between
layers gets incoherent the better the sample quality -
characterized by the full width at half maximum (FWHM) of the
rocking curve - and suggests that the coupling between planes is
much less than the commonly assumed 0.3~eV \cite{kempa3}.

In the quantum limit (QL) when only the lowest Landau levels of
graphite are occupied ($B > 2~$T), the longitudinal electrical
resistivity shows a reentrance to a metallic-like state below a
field-dependent temperature $T_m(B)$ \cite{yakov1,yakovprl}. The
function $T_m(B)$ depends on the dimensionality of the graphite
sample and it oscillates as a function of field for quasi 2D
samples. This behavior might be an evidence for field-induced
superconductivity at the QL in graphite, discussed in
Refs.~\cite{man,tes}. In the QL the Hall resistance $R_H$ shows
clear signs of the quantum Hall effect for samples with small FWHM
\cite{yakovprl}.

For ideal graphite the conduction electrons are expected to follow
a Dirac dispersion relation. These quasiparticles (QP) should have
some similarities with the nodal QP of the $d$-density wave (DDW)
state, which applies also to high-temperature superconductors. For
a not superconducting DDW QP the violation of the Wiedemann-Franz
(WF) law \cite{wf} has been predicted \cite{yang} as a
characteristic of the relativistic spectrum of the QP. Briefly, in
the DDW state the QP can carry electrical current much more
effectively than thermal current in the limit of very small QP
density, a situation that applies to graphite (Fermi Energy $E_F
\sim 200~$K). The theoretical work of Ref.~\cite{sha}, however,
goes a step further taking into account the presence of a magnetic
field within the DDW state in the calculations. This work finds
that the WF law is restored at low enough temperatures (compared
with $E_F$) independently of the applied field. We note that in
none of these references quantum Hall states have been taken into
account in the calculations. Taking into account recently
published experimental evidence \cite{yakovprl}, these theories
cannot be applied straightforwardly to graphite. The aim of our
work is to check whether the WF law applies in graphite in
particular in the QL regime.

 The magnetic field dependence of the thermal
conductivity of strongly anisotropic HOPG samples was already measured in
the past but one does not find curves  in the published literature with
the necessary resolution in the QL regime. There are at least two
unpublished studies which deserve our attention. In his Ph.D. thesis,
Ayache \cite{aya} measured $\kappa(T,B)$ and observed well defined
oscillations of $\kappa(B)$ at fields above 2~T and at constant
temperatures below 10~K. This behavior was apparently also observed by
Woollam \cite{woo} but the curves were not included in the corresponding
paper. Interestingly, both authors emphasized that the oscillations in
$\kappa(B>2~$T) as a function of field were apparently in phase with the
electrical resistivity oscillations, although the electronic contribution
to $\kappa$ at high fields freezes out according to the Wiedemann-Franz
law. In spite of this striking result neither the curves nor any analysis
of the data was published to our knowledge.

High-resolution measurements of the field dependence of the
thermal conductivity of quasi 2D HOPG samples have nowadays
special relevance. In particular because evidence for a
quantum-Hall behavior in 2D samples in the QL has been recently
reported \cite{yakovprl}. The behavior of $\kappa(T,B)$ in
graphite is not only important to understand the nature of the QP
in this material at the QL but also gives us the chance to study
the thermal transport in the quantum-Hall effect (QHE) regime. We
note that the thermal transport in the QHE regime still remains an
unclear problem since the discovery of this effect. In this work
we have measured the longitudinal thermal conductivity $\kappa$ of
a well characterized HOPG sample as a function of magnetic field
applied parallel to the $c-$axis of the graphite structure. These
measurements were accompanied by measurements of the longitudinal
and Hall resistances. The results clearly show that the WF-law is
violated in the QL, whereas deviations are observed at lower
fields.

\section{Experimental Details}

The HOPG sample from the Union Carbide company measured in this work was
selected because of its clear quasi-2D properties observed in the
electrical and Hall resistivity. This behavior is correlated with the
small FWHM$=0.26^\circ$ of the rocking curve, which is a measure of the
misorientation relative to the $c-$axis of the crystallites in the sample.
The measured FWHM is one of the smallest we have obtained for HOPG
samples. We note that most of the graphite samples studied in the
literature of the 70's and 80's had much larger FWHM. Our experimental
evidence indicates that in samples with FWHM values larger than $\sim
0.5^\circ$ the transport properties show clear sign of 3D behavior and
coherent transport in the $c-$direction\cite{kempa2,kempa3} and therefore
they do not reflect true 2D behavior of ideal graphite. The dimension of
our sample was: length = 1.6~mm, width = 1.2~mm and thickness $ \simeq
60~\mu$m. The room temperature out-of-plane/basal-plane resistivity ratio
at $B = 0$ of the sample was $\rho_c/\rho_b  \sim 5 \times 10^4$.
Furthermore, this sample does not show any maximum in the $c-$axis
resistivity as a function of angle for fields parallel to the graphene
planes \cite{kempa3}. This indicates incoherent electrical transport
expected for weak-coupled conducting planes and for samples with a low
density of defects.

The Hall resistance was measured using the van der Pauw configuration with
a cyclic transposition of current and voltage leads \cite{kope4,lev} at
fixed applied-field polarity as well as magnetic field reversal; no
difference in $R_h(H,T)$ obtained with these two methods was found. For
the measurements, silver past electrodes were placed on the sample
surface, while the resistivity values were obtained in a geometry with an
uniform current distribution through the sample cross section. All
resistance measurements were performed in the Ohmic regime. The absolute
value of the longitudinal (basal-plane) resistivity at zero field was
$\rho_b(6~$K$,0) \simeq 2.4 \mu\Omega$cm and $2.6 \mu\Omega$cm at 10~K.
The error in the absolute value is estimated to be $\sim 30\% $ due to
geometrical errors. At 10~K $\rho_b$ reaches its low-temperature rest
value within 10\%. The resistivity increases by two orders of magnitude
for an applied field of 1~T at $T \le 10~$K.

 For the longitudinal thermal conductivity measurement the temperature gradient
(of the order of 200 to 300 mK) was measured using a previously field- and
temperature-calibrated type E thermocouples with a dc-picovoltmeter
\cite{iny}.  The thermocouple ends were positioned one at the top and the
other at the bottom of the main surface of the sample. A detailed
calibration below 8~K was performed because in this temperature region the
thermopower of our thermocouple is specially sensitive on the magnetic
field with a non-monotonous dependence\cite{iny}. The experimental
arrangement was recently used to study the longitudinal and Hall thermal
conductivities of high-temperature superconducting
crystals\cite{oca1,oca2}. We note that in general the measured thermal
conductivity is $\kappa = \kappa_i - TS^2\sigma_i$, where $\kappa_i$ is
the ``real" thermal conductivity of the sample, $S$ the thermopower and
$\sigma_i$ the electrical conductivity. In the case of our graphite sample
the correction term to $\kappa$ is four orders of magnitude smaller than
$\kappa_i$ at 10~K.

Our system enables us to measure $\kappa(B)$ with a relative resolution
better than $0.1\%$ above 5~K. The thermal stability was better than 10~mK
in the whole temperature range 5~K~$\le T \le 20~$K and magnetic field
0~T$ \le B \le 9~$T. The absolute error in the thermal conductivity was
estimated to be $\le 30\%$. The obtained absolute value of $\kappa$ and
its temperature dependence are similar to those from previous
studies\cite{aya,woo,her,iss,dre}. For example, at 10~K we obtain
$\kappa(0) \simeq 130~$W/mK and $\kappa(0) \simeq 33~$W/mK at 5~K and zero
fields. In all magnetic-field runs $\kappa(B)$ showed reversible behavior.
Irreversible behavior has been observed but we could prove that it was
related to small temperature drifts since in the temperature range of the
measurements $\kappa$ depends strongly on temperature.

\section{Results and Discussion}

\begin{figure}
\epsfig{file=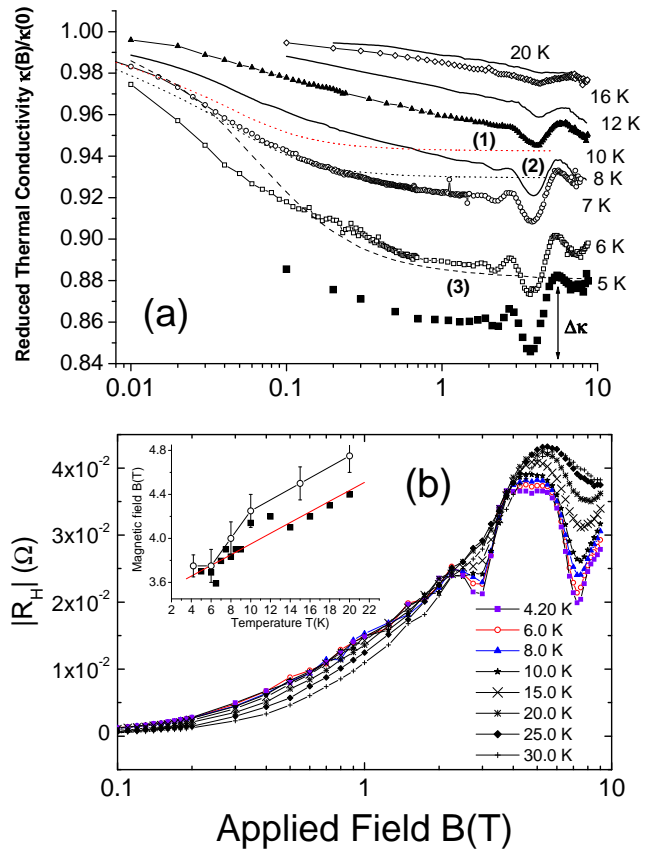,width=\columnwidth} \caption{(a) Reduced thermal
conductivity $\kappa(B)/\kappa(0)$ vs. applied magnetic field at different
constant temperatures. The lines (2) and (3) were calculated from the
Wiedemann-Franz law Eq.~(\protect\ref{wf}) and Eq.~(\protect\ref{3}) at
10~K and 6~K using the measured field dependent electrical resistivity for
the same sample. Curve (1) was obtained with the measured resistivity at
10~K but with a lower Lorenz number $L = 2.0 \times
10^{-8}~$W$\Omega$K$^{-2}$. (b) Absolute value of the Hall resistance as a
function of applied field. The inset shows the temperature dependence of
the field at the onset of the plateau in $R_H$ at $\sim 3.7~$T ($\circ)$
and the position of the minimum in $\kappa$ ($\blacksquare)$.} \label{1}
\end{figure}

\begin{figure}
\epsfig{file=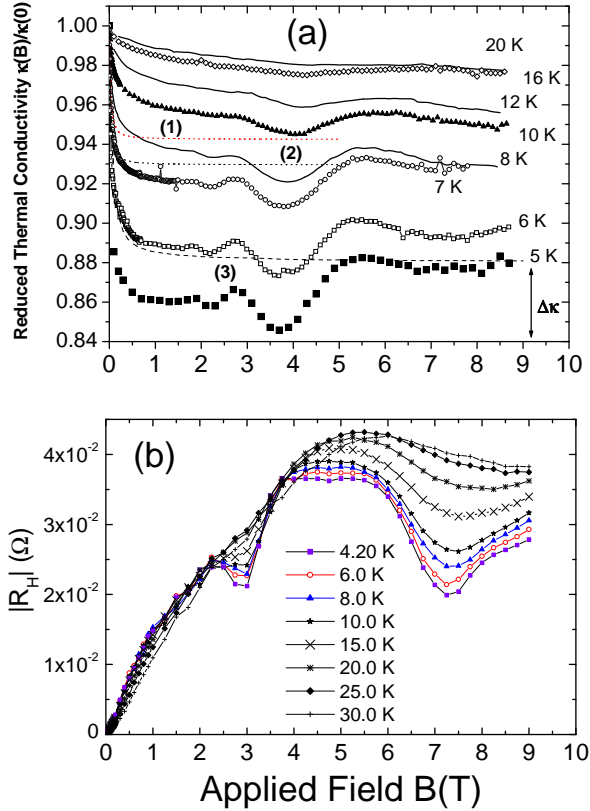,width=\columnwidth} \caption{The same data as in
Fig.~1 but plotted in a linear field scale.} \label{11}
\end{figure}

Figure~\ref{1}(a) shows the reduced longitudinal thermal conductivity
$\kappa(T,B)/\kappa(T,0)$ as a function of the field applied parallel to
the $c-$axis at different constant temperatures between $5~$K to 20~K. In
Fig.~\ref{1}(b) we show the Hall resistance of the same sample. The
decrease of $\kappa$ with field can be related to the decrease of the
electronic contribution. The clear oscillations in $\kappa(B)$ observed at
$B > 1~$T are apparently due to the quantization of the Landau levels and
the crossing of the Fermi energy as the oscillations in the Hall effect
indicate, see Fig.~\ref{1}(b). Figure~\ref{11} shows the same data as
Fig.~\ref{1} but in a linear field scale.

It is noticeable the appearance of the plateau-like features at $B
\sim 2~$T and 4~T in the Hall resistivity that clearly suggests
the occurrence of the Quantum Hall effect (QHE) in graphite. In
fact, the temperature dependence of the maximum slope $({\rm
d}|R_H|/{\rm d}B)_{\rm max}$ vs. $T^{-1}$ between two plateaus at
3.5~T, and measured to 70 mK shows a temperature dependence
$T^{-k}$ with an exponent $k = 0.42$ similar to that found in QHE
systems \cite{yakovprl}. This result is not unexpected taking into
account the quasi 2D structure of the sample. We note that the
occurrence of the integral QHE in graphene has been predicted
recently \cite{zhe}. The reason why it was not found before is
related to the sample quality which affects the 2D behavior of the
transport properties. Our studies show that the dimensionality is
strongly affected by internal lattice defects, some of them appear
to short circuit the graphene planes. The fact that we want to
stress is the good correspondence between the features measured in
$\kappa$ and $R_H$ at the QL. To recognize this we show in the
inset of Fig.~\ref{1}(b) the temperature dependence of the field
at the onset of the plateau in $R_H$ at $\sim 3.7~$T ($\circ)$ and
the position of the minimum in $\kappa$ ($\blacksquare)$.

\begin{figure}
\epsfig{file=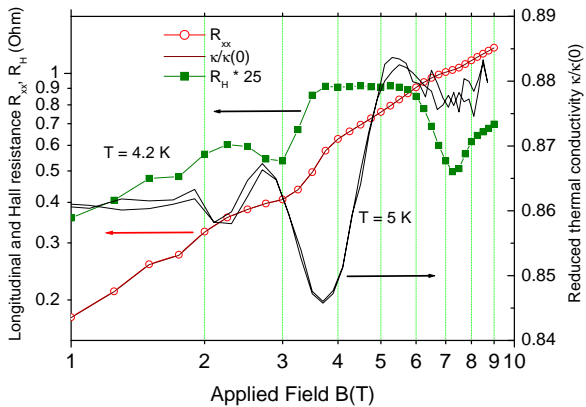,width=\columnwidth} \caption{Longitudinal and Hall
resistances (left scale) and the reduced thermal conductivity (right
scale) as a function of magnetic field applied parallel to the $c-$axis.
The two measured lines of the thermal conductivity data were taken
increasing and decreasing field. Within the error there are no significant
hysteresis.} \label{2}
\end{figure}

Figure~\ref{2} shows in more detail the field dependence of the
basal-plane and Hall resistances, taken at 4.2~K, as well as of the
thermal conductivity at 5~K. The thermal conductivity data shown in this
figure were taken increasing and decreasing field. As seen in the figure,
no significant hysteresis is observed. This result is in contrast to the
hysteresis observed by Ayache in his Ph.D. work \cite{aya}. We speculate
that a small temperature drift might have been the reason for the observed
hysteresis.

Can we understand the decrease with field and the oscillations observed in
$\kappa(B)$ within the Wiedemann-Franz relation? To answer this question
we proceed as follows. We assume that the thermal transport of graphite is
given by two contributions \cite{kel}:
\begin{equation}
\kappa= \kappa_p(T) + \kappa_e(T,B)\,, \label{kappa}
\end{equation}
where $\kappa_p$ is due to the phonons, the contribution of the
atomic lattice with the appropriate lattice anisotropy, and
$\kappa_e$ due to conduction electrons. Usually one assumes that
the field dependence of the thermal conductivity is given only by
the electronic part $\kappa_e(T,B)$, which can be estimated with
the WF relation. This universal relation relates the electrical
resistivity $\rho(T,B)$ with the thermal conductivity due to
electrons by
\begin{equation}
\frac{\kappa_e \rho}{T} = L_0 \label{wf}\,,
\end{equation}
 through the universal
constant $L_0 = 2.45 \times 10^{-8}~$W$\Omega$K$^{-2}$ at low
enough temperatures. One can recognize easily the difficulty to
measure accurately the field dependence of the electronic
contribution to the thermal transport above $\sim 4~$K due to the
small electronic contribution. From Eq.~(\ref{wf}) we expect for
well ordered HOPG samples at zero field a ratio between the
electronic and total thermal conductivity $\kappa_e/\kappa < 0.15$
at 5~K, and $< 0.05$ at 10~K. Literature data are in general in
good agreement with these estimates\cite{aya,her}.

 The relation (\ref{wf})
holds strictly for elastic or quasielastic electron scattering and
therefore the range of validity is usually set, either at low enough
temperatures where the resistivity is temperature independent (impurity
scattering dominates), or at high enough temperatures where the
electron-phonon scattering is large \cite{am}. For the sample measured in
this work, the temperature dependence of the electrical resistivity
indicates a saturation below 10~K (curves for a similar sample can be seen
in Refs.~\cite{yakovadv,ser}) and therefore at $T \le 10~$K we expect to
be roughly in the validity range of the WF-law. From the measured field
dependence of the electrical resistivity we can calculate the relative
change of the total thermal conductivity at a fixed temperature as:
\begin{equation}
\frac{\kappa(B)}{\kappa(0)} = \frac{\kappa_e(B) -
\kappa_e(0)}{\kappa(0)} + 1\,, \label{red}
\end{equation}
  assuming that the
phonon conductivity does not depend on magnetic field. In
Fig.~\ref{1}(a) we show three curves calculated with
Eqs.~(\ref{red}) and (\ref{wf}) using the measured $\rho(B)$ at
10~K (curves (1) and (2)) and 6~K (3). Curve (1) was obtained with
the same parameters as (2) but with $L_0 = 2.0 \times
10^{-8}$W$\Omega$K$^{-2}$, assuming a decrease of $L_0$ due to the
possible influence of the inelastic scattering.

From the comparison between the computed curves and the experimental ones
one would tend to conclude that Eqs. (2) and (3) provide reasonably well
the overall decrease of $\kappa$ with field, within the geometrical errors
in the measurement of both conductivities. Nevertheless we should note
that the WF law and Eq.~(\ref{red}) do not reproduce accurately the
measured field dependence, see Fig.~1(a). Due to the electrical
resistivity increase (a factor $\sim 100$ from zero field to 1~T at $T \le
10~$K) the electronic contribution, according to the WF law, becomes
negligible. Therefore the electronic contribution to $\kappa$ should be
negligible,
 e.g. at $B > 0.4~$T and $T = 10~$K, and a saturation of $\kappa$ at larger
fields is expected. This is not observed experimentally. This means also
that the relatively small oscillations observed in the longitudinal
resistivity above 1~T that accompany the features in the Hall effect (see
Fig.2) should not affect $\kappa$ according to Eq.~(\ref{wf}) in contrast
to the experimental results, see Fig.1(a). These results clearly indicate
that the WF law in its original form fails to explain the field dependence
of the thermal conductivity in the QL of graphite and in the measured
temperature range.

\begin{figure}
\epsfig{file=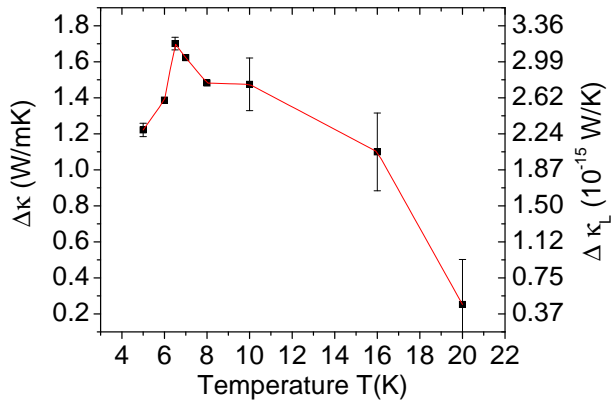,width=\columnwidth} \caption{The thermal
conductivity difference between the minimum (at $B \simeq 3.7~$T)
and maximum (at $B \simeq 5.5~$T) $\Delta k$ (see
Fig.~\protect\ref{1}) as a function of temperature. The right
scale provides the estimate values per graphene layer (within a
geometrical error of $\sim 30$\%).} \label{3}
\end{figure}

For any future comparison with theory it may be useful to provide the
absolute value of the oscillation amplitude between minimum and maximum as
defined in Fig.~\ref{1}(a). Figure~\ref{3} shows the experimental $\Delta
\kappa$ obtained from Fig.~\ref{1}(a) or Fig.~\ref{11}(a). It shows a
tendency to saturation around $\Delta \kappa \sim 1.5~$W/mK for $T <
10~$K. The oscillation amplitude per graphene layer can be estimated from
$\Delta\kappa_L \sim \Delta\kappa a / N$, where $a$ is the distance
between layers and $N$ the number of layers measured in parallel in our
experiment within the $60~\mu$m thickness of the sample. Using this
approximation we obtain $\Delta\kappa_L \sim 3 \times 10^{-15}~$W/K for $T
< 10~$K.

In the following we discuss our results taking into account
relevant work. A recently published theoretical work calculated
the thermal conductivity of a 2D electron gas at low temperatures
and in a quantizing magnetic field \cite{kar}. Although some of
the assumptions done there may not be valid for graphite, it is
interesting to note that this work shows that the WF-law is
violated for small Landau level broadening and at low enough
temperatures when the diffusion mechanism dominates. According to
this work the deviations from the WF law are due to the energy
derivatives of the longitudinal electrical conductivity. At low
enough temperatures and small level broadening the numerical
results show a ``two-peak" behavior, i.e. $\kappa$ as a function
of magnetic field shows two maxima in the field region between two
neighboring plateaus in the Hall resistance. A similar result has
been obtained in previous theoretical work \cite{oji,bla}. In our
case, however, we obtain the striking result that $\kappa_e(B)$
{\em decreases} in this field region whereas {\em increases} with
field in the plateau region and reaches a maximum at the end of
the corresponding plateau, see Fig.~\ref{2}. If we would have
localized QP in the field region of the plateau we would naively
expect a decrease of $\kappa_e$. On the other hand, the opposite
behavior may be also possible, i.e. an increase of $\kappa_e$ with
field, if the density of interacting QP would decrease in this
field region and the main scattering mechanism is given by QP-QP
scattering, as in the case of high-temperature superconductors
\cite{yu2}. In this case the theoretical description of the
thermal conductivity may become more complicated to handle.

According to recently published theory \cite{khv,GorbarPRB2002}, a
magnetic field applied perpendicular to the graphene planes opens
an insulating gap in the spectrum of Dirac fermions, associated
with an electron-hole pairing, leading to the excitonic insulator
state below a field dependent transition temperature. The
experimental value of the critical field of the field-driven
metal-insulator transition in graphite is $\sim$50 times smaller
than the predicted in Refs.~\cite{GorbarPRB2002}. The discrepancy
can be understood, however, assuming that the Coulomb coupling,
given by the dimensionless parameter $g = 2\pi e^2 /\epsilon_0 v$
($\epsilon_0$ is the dielectric constant)
\cite{khv,GorbarPRB2002},
 drives the system very close to the excitonic instability. In
 this case, the threshold field $B_c$ can
be well below the estimated value of $2.5~$T. The above analysis,
together with the experimental evidence that only the
perpendicular component of the applied field drives the MIT
\cite{kempa3}, appear to support the theoretical expectations of a
field-induced excitonic insulator state in graphite. In this case
and according to Ref.~\cite{fer} we would expect a monotonic
decrease of $\kappa_e(B)$ with field and a kink at $B \sim B_c$
with a plateau region in the insulating-like state of the QP at $B
> B_c$. In the temperature range of our experiments the results do
not show a clear kink at $B \sim 0.1~$T nor a plateau at higher fields and
$T \ge 6~$K, see Fig.~\ref{1}(a), although one may tend to recognize it at
$T = 6~$K and $B \sim 0.5~$T when the data are plotted in a linear field
scale, see Fig.~\ref{11}(a). At the temperature limit of our system (5~K)
the density of points at $B < 1~$T is too low to assure the existence or
non-existence of a kink. Measurements at lower temperatures are needed to
enhance the relative contribution of the QP to $\kappa$ and check whether
the predicted feature is observable. This issue will be studied in the
future.

Can the oscillations in $\kappa(B > 1~$T) be due to the lattice
contribution $\kappa_p(B)$ via electron-phonon interaction as, for
example, in antimony \cite{long}? The temperature dependence of $\kappa
\propto T^{2.4}$ at 5~K~$< T < 30~$K speaks for phonon scattering by grain
boundaries and not by electrons \cite{kel}. The inelasticity parameter
$\eta = v /\lambda \omega_c$ (here $v$ is the sound velocity, $\lambda$
the magnetic length and $\omega_c$ the cyclotron frequency), that provides
an estimate of the efficiency of the electron-phonon scattering, is $\sim
0.01$ at 4~T for graphite. Therefore, unless there is an intersection of
Landau levels that favors acoustic phonon transitions \cite{gol}, it does
not seem that the phonon-electron scattering can be significantly enhanced
at high fields in graphite. The overall correlation of $\kappa(T,B)$ with
the Hall resistivity indicates that the origin of the oscillations in
$\kappa(B)$ should be related to a pure QP phenomenon.

In summary, high-resolution measurements of the magnetic field dependence
of the thermal conductivity in a quasi 2D sample of graphite show clear
oscillations in the quantum limit. The Hall effect for the same sample
shows quantum Hall effect features which are correlated to the features
observed in $\kappa$. With the measured longitudinal electrical
resistivity we show that the observed oscillations in $\kappa$ cannot be
explained with the original WF law. Lower temperature measurements as well
as an appropriate theoretical framework for graphite are highly desirable.

\begin{acknowledgements}
We thank K. Ulrich for his help with the calibration of the
thermocouple, M. A. Vozmediano and F. Guinea for fruitful
discussions and their interest on this work, and E. J. Ferrer and
V. P. Gusynin for clarifying details of their theoretical work.
This research was partially supported by the DFG under DFG ES
86/6-3, FAPESP and CNPq.
\end{acknowledgements}


\begin{thebibliography}{99}
\bibitem{kel}B. T. Kelly,  {\it Physics of Graphite}, p. 267 ff,
p. 293, p. 322 ff. Applied
Science Publishers LTD, London and New Jersey (1981).

\bibitem{dre}M. S. Dresselhaus, G. Dresselhaus, K. Sugihara, I. Spain,
and H. A. Goldberg, in
 {\it Graphite Fibers and Filaments},
Springer Series in Material Science Vol. 5, pp. 179-188
(Springer-Verlag, 1989).

\bibitem{dil}R. O. Dillon, and I. L. Spain, J. A. Woollam, and W. H. Lowrey,
J. Phys. Chem. Solids
 {\bf 39}, 907-922 (1978). Ibid,  {\bf 39}, 923-932 (1978).

\bibitem{kempa1}H. Kempa, Y. Kopelevich, F. Mrowka,
A. Setzer, J. H. S. Torres, R. H\"ohne, and P. Esquinazi,
Solid State Commun. {\bf 115}, 539 (2000).

\bibitem{kempa2} H. Kempa, P. Esquinazi, Y. Kopelevich,
Phys. Rev. B {\bf 65}, 241101(R) (2002).

\bibitem{yakovadv}Y. Kopelevich,  P. Esquinazi, J. H. S. Torres, R. R. da
Silva, and H. Kempa, Advances in Solid State Physics {\bf 43}, in press
(2003).

\bibitem{khv}D. V. Khveshchenko,  Phys. Rev. Lett. {\bf 87},
206401 (2001); {\it ibid} {\bf 87}, 246802 (2001).

\bibitem{das}D. Das and S. Doniach, Phys. Rev. B {\bf60}, 1261 (1999);
Phys. Rev. B {\bf64}, 134511 (2001).

\bibitem{kempa3}H. Kempa, H. C. Semmelhack, P. Esquinazi and Y.
Kopelevich, Solid State Commun. {\bf 125}, 1 (2003).

\bibitem{yakov1} Y. Kopelevich, V. V. Lemanov, S. Moehlecke,
J. H. S. Torres, Phys. Solid State {\bf 41}, 1959 (1999).

\bibitem{yakovprl} Y. Kopelevich,  J. H. S. Torres, R. R. da Silva, F. Mrowka, H. Kempa and P.
Esquinazi, Phys. Rev. Lett. {\bf 90}, 156402 (2003)

\bibitem{man} T. Maniv, V. Zhuravlev, I. Vagner, and P. Wyder, Rev.
Mod. Phys. {\bf 73}, 867 (2001).

\bibitem{tes} Z. Tesanovic and M. Rasolt, Phys. Rev. B {\bf 39}, 2718 (1989).

\bibitem{wf}G. Wiedemann and R. Franz, Ann. Phys. {\bf 89} (1853).

\bibitem{yang}X. Yang and C. Nayak, Phys. Rev. B {\bf 65}, 064523 (2002).

\bibitem{sha}S. G. Sharapov, V. P. Gusynin and H. Beck, Phys. Rev. B {\bf
67}, 144509 (2003).

\bibitem{aya}C. Ayache, Ph.D. Thesis, Grenoble 1978 (unpublished). One of
the curves of the thesis can be found in {\it Landolt-B\"ornstein}, New
Series III/15c, page 433 (1991).

\bibitem{woo}J. A. Woollam, Phys. Rev. B {\bf 3}, 1148 (1971).

\bibitem{kope4}Y. Kopelevich, V. V. Lemanov, S. Moehlecke, and J. H. S. Torres,
Sov. Phys. Solid State {\bf 26}, 1607 (1984).

\bibitem{lev}M. Levy and M. P. Sarachik, Rev. Sci. Instrum. {\bf 60}, 1342
(1989).

\bibitem{iny} A. Inyushkin,  K. Leicht, and P. Esquinazi, Cryogenics {\bf 38}, 299 (1998).

\bibitem{oca1}R. {Oca\~{n}a} and  P. Esquinazi, Phys. Rev. Lett. {\bf 87},
167006 (2001).
\bibitem{oca2}R. {Oca\~{n}a} and  P. Esquinazi, Phys. Rev. B {\bf 66},
064525 (2002).

\bibitem{her}J. Heremans, M. Shayegan, M. S. Dresselhaus, and J.-P- Issi,
Phys. Rev. B {\bf 26}, 3338 (1982).

\bibitem{iss}J.-P- Issi and J. Heremans, and M. S. Dresselhaus, Phys. Rev.
B {\bf 27}, 1332 (1983).

\bibitem{zhe}Y. Zheng and T. Ando, Phys. Rev. B {\bf 65}, 245420 (2002).

\bibitem{am}See, for example, N. W. Ashcroft and N. D. Mermin, in {\it Solid State Physics},
Holt-Saunders International Editions, 1976.

\bibitem{ser} M. S. Sercheli, Y. Kopelevich, R. Ricardo da Silva, J. H. S. Torres, and C. Rettori,
Solid State Commun. {\bf 121}, 579 (2002).

\bibitem{kar} V. C. Karavolas and G. P. Triberis, Phys. Rev. B
{\bf 59}, 7590 (1999).

\bibitem{oji} H. Oji, Phys. Rev. B {\bf 29}, 3148 (1984); J. Phys.
C: Solid State Phys. {\bf 17}, 3509 (1984).

\bibitem{bla}Ya. M. Blanter, D. V. Livanov, and M. O. Rodin, J.
Phys.: Condes. Matter {\bf 6}, 1739 (1994).

\bibitem{yu2}R.C. Yu, M. B. Salamon, Jian Ping Lu, and W. C. Lee, Phys. Rev. Lett. {\bf 69},
1431 (1992).

\bibitem{GorbarPRB2002} E. V. Gorbar, V. P. Gusynin, V. A. Miransky, and
I. A. Shovkovy, Phys. Rev. B {\bf 66}, 045108 (2002).

\bibitem{fer} E.J. Ferrer, V.P. Gusynin, V. de la Incera,
cond-mat/0203217.

\bibitem{long}J. R. Long, C. G. Grenier, and J. M. Reynolds, Phys.
Rev. B {\bf 140}, A187 (1965).

\bibitem{gol}V. N. Golovach and M. E. Portnoi, cond-mat/0202179.


\end{thebibliography}
\end{document}